# On the Tachocline Zone Location in the Sun, the Luminosity Transport time scale, the Rotational Inertia and their Time Variation in Standard Solar Evolution Models

Mandyam N Anandaram*

**Abstract**

Studies with GONG Standard Solar Evolution Models sampling the evolution of the sun from its ZAMS stage show the following. The location of the tachocline zone is nearly fixed as it is not affected by shell burning although it co-moves with the expansion of the sun up to the present age of 4.6 Gyr. The luminosity transport time scale of the sun is entirely dominated by photon diffusion and during the evolution has decreased from over 204000 years to 187000 years. The rotational inertia of the sun shows a small gradual increase from $\mathbf{6.73 \times 10^{46}}$ kgm² to $\mathbf{7.09 \times 10^{46}}$ **kgm²** at present but the inertia constant decreases from 0.0911 to 0.0736 during the same period.

## 1. Introduction

The Model S (also labeled No.24) is a standard solar evolution model (SSEM) evolved to the estimated age of 4.6 Gyr as part of the

* Bangalore University, Bangalore;   mnanandaram@gmail.com

                                                                    

GONG project [1]. The GONG group applied the methods of helioseismic analysis to this high resolution model which lists data over 2482 grid points (concentric spherical shells) to determine the location of the tachocline zone (TZ) at the bottom of the convective zone (CZ). In Model S (SSEM No.24) this location was determined by [1] as corresponding to a radius fraction of 0.7112 and a mass fraction of 0.9753 of the model. A study of the dependence of the TZ location on the age of the solar evolution has been made by studying similar SSEMs of the GONG project from zero age main sequence (ZAMS) stage onwards. This study is done (section 2) on 15 GONG SSEMs [2] and one YREC SSEM No.741 [3] of nearly same age as Model S. As the determination of the TZ location implied that the extent of the solar radiative zone (RZ) is also known in all the SSEMs, it was decided to investigate the variation with model age of the characteristic luminosity transport timescale (section 3) and the rotational moment of inertia (section 4) as the required opacity and density data were available for all model grid points.

A remark on the nature of the tachocline zone is relevant here. This region is also referred to in literature as the interface layer, or overshoot zone. It is known [4] that the bulk of the radiative interior (RZ) slowly rotates almost like a rigid body and it ends in a region of radial shear known as tachocline. This shear is due to the latitude dependent differential rotation [5] in the entire convective zone. The rotation rate of the RZ is known to be nearly equal to the rotation rate of middle latitudes on the sun which corresponds to a value midway between the slow polar rate and the faster equatorial rate. This rotation frequency is about $4.33 \times 10^{-7} Hz$ corresponding to a sidereal rotation period of about 26.7 days. This value will be used to estimate the angular momentum of the sun in section 4.

## 2. Study of the Tachocline Zone (TZ) location in SSEMs

The TZ is a thin layer below the bottom of convective zone (CZ) in the sun where the solar radiative zone (RZ) ends. The bottom shell of the CZ is identified by the constancy of the hydrogen mass fraction X throughout the CZ and hence the previous shell is taken as the location of the TZ. These details may be seen in Figure 1 which plots Model S (SSEM #24) and in Figure 2 which plots zero age model (ZAMS, SSEM #01) of the sun. The upper parts of both figures show normalized temperature, density, pressure, luminosity, radius, X and Y as functions of the mass fraction whereas the lower parts show them as functions of the radius fraction. The vertical blue dotted line marks the location of the tachocline zone (TZ) which separates the inner radiative zone (RZ) containing about 97.5 % of mass at left from the outer convective zone (CZ) containing 2.5% mass at right (see Table 1). The step up in X and step down in Y may be seen at TZ in Figure 1. This stepping is absent in the ZAMS model of the sun shown in Figure 2 since the chemical composition is uniform at zero age. The grid point number of the TZ location was determined from the model data after drawing normalized plots similar to Figure 1 for all the bottom of the CZ.

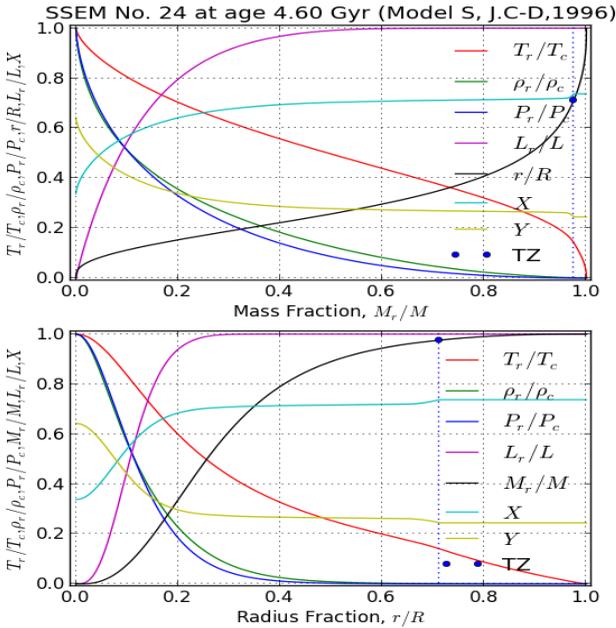

Fig. 1. The above graphs show all the normalized parameters of the Model S (SSEM #24).

Table 1. The first three columns list the basic data of the 16 SSEMs. The next four columns show the location of the tachocline zone (TZ) as shell number, its radius fraction depth $f_{TZ}$, Mass fraction depth, X in the convective zone (CZ) and the buoyant diffusion contribution dX at CZ.

| SSEM No. | Age Gyr | R km | Shell No. | $f_{TZ}$ $(r/R)_{TZ}$ | $M_r/M$ $(M_r/M)_{TZ}$ | X $X_{CZ}$ | dX $X_{CZ} - X_{01}$ |
|---|---|---|---|---|---|---|---|
| 01 | 0.000 | 609615 | 990 | 0.721728 | 0.969772 | 0.7091 | 0.0000 |
| 03 | 0.051 | 611212 | 950 | 0.720744 | 0.969454 | 0.7094 | 0.0003 |
| 05 | 0.185 | 614395 | 951 | 0.720486 | 0.969517 | 0.7102 | 0.0011 |
| 07 | 0.386 | 617917 | 952 | 0.719594 | 0.969618 | 0.7115 | 0.0024 |
| 09 | 0.660 | 622193 | 954 | 0.718829 | 0.969921 | 0.7132 | 0.0041 |
| 11 | 1.010 | 627537 | 957 | 0.718112 | 0.970396 | 0.7153 | 0.0062 |
| 13 | 1.442 | 634239 | 961 | 0.717352 | 0.971018 | 0.7180 | 0.0089 |
| 15 | 1.959 | 642636 | 966 | 0.716418 | 0.971757 | 0.7212 | 0.0121 |
| 17 | 2.565 | 653172 | 973 | 0.715970 | 0.972769 | 0.7250 | 0.0159 |
| 18 | 2.903 | 659430 | 976 | 0.714841 | 0.973125 | 0.7269 | 0.0178 |
| 20 | 3.603 | 673385 | 984 | 0.713478 | 0.974067 | 0.7312 | 0.0221 |
| 21 | 3.919 | 680171 | 988 | 0.712921 | 0.974491 | 0.7331 | 0.0240 |
| 22 | 4.213 | 686811 | 992 | 0.712501 | 0.974898 | 0.7348 | 0.0257 |
| 23 | 4.407 | 691330 | 995 | 0.712039 | 0.975118 | 0.7360 | 0.0269 |
| 24 | 4.600 | 695990 | 997 | 0.711177 | 0.975251 | 0.7370 | 0.0279 |
| 741 | 4.550 | 695978 | 576 | 0.711115 | 0.975201 | 0.7362 | 0.0309 |

16 SSEMs. This TZ grid point number is listed in the fourth column of Table 1 along with the corresponding values of the radius fraction, mass fraction and X for all the SSEMs studied. Only in the case of the ZAMS model (SSEM #01) in which *X* is uniform the local polytropic index was also computed and plotted to identify the step up in *X* and the step down in Y (helium) and Z (metals) at TZ in all the SSEMs is shown in Figure 3 with more detail for *X* in the graph at top in which a black ×-mark locates TZ at the base of the CZ. In the left half of Figure 3, the top down order of the graphs for *X* starts with age zero at top left and end at present age below.

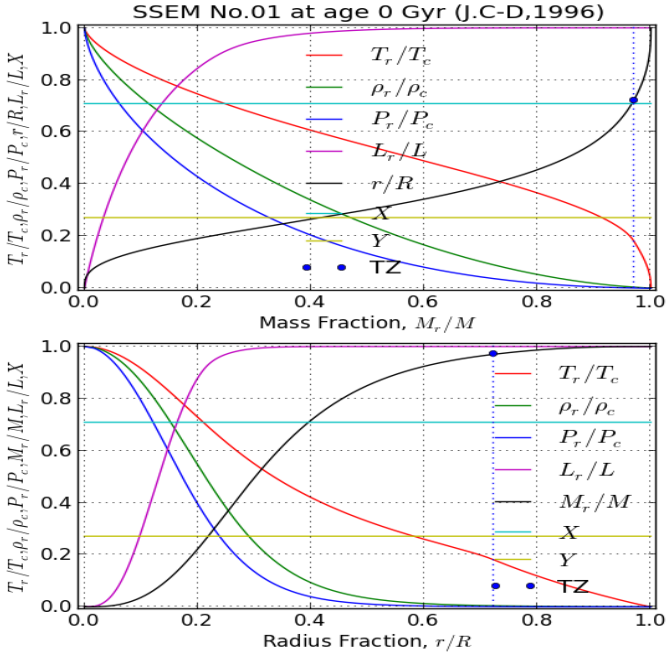

Fig. 2. The graphs show all the normalized parameters of the ZAMS Model (SSEM #01).

In the right half notice that this order is completely reversed at TZ. In the middle and bottom graphs, **h**elium (Y) and metals (Z) profiles are drawn for 15 GONG SSEMs. In the CZ at right which extends beyond the TZ at about 0.71% of r/R the buoyant diffusion is continuously increasing the initial ZAMS stage value of X but the continued settling into the core is depleting the levels of Y and Z. In the core at left of Figure 3, it is seen that X decreases while Y and Z increase with age.

The determination of the TZ location as a grid number implies an inherent error of at least one shell width of about 600 kilometers. This means that the width of the solar radiative zone (RZ) is also known to a good accuracy and its variation as a total fraction of model radius is negligible throughout the 4.6 Gyr period of solar evolution. This width is listed in the 5th column of Table 1 while the mass fraction up to this point is listed in the 6th column.

53

However due to the expansion caused by shell burning the solar radius is slowly increasing with age and hence the TZ location is co-moving outwards. But the TZ location as a fraction of the solar radius appears to contract by a negligible amount. This can be clearly seen in Figure 3. The mean value of the mass fraction is thus about 97.3% at a radius fraction of 71.7%. This observation contradicts the basic premise of Eddington's polytropic solar model that the entire sun is in radiative equilibrium [17].

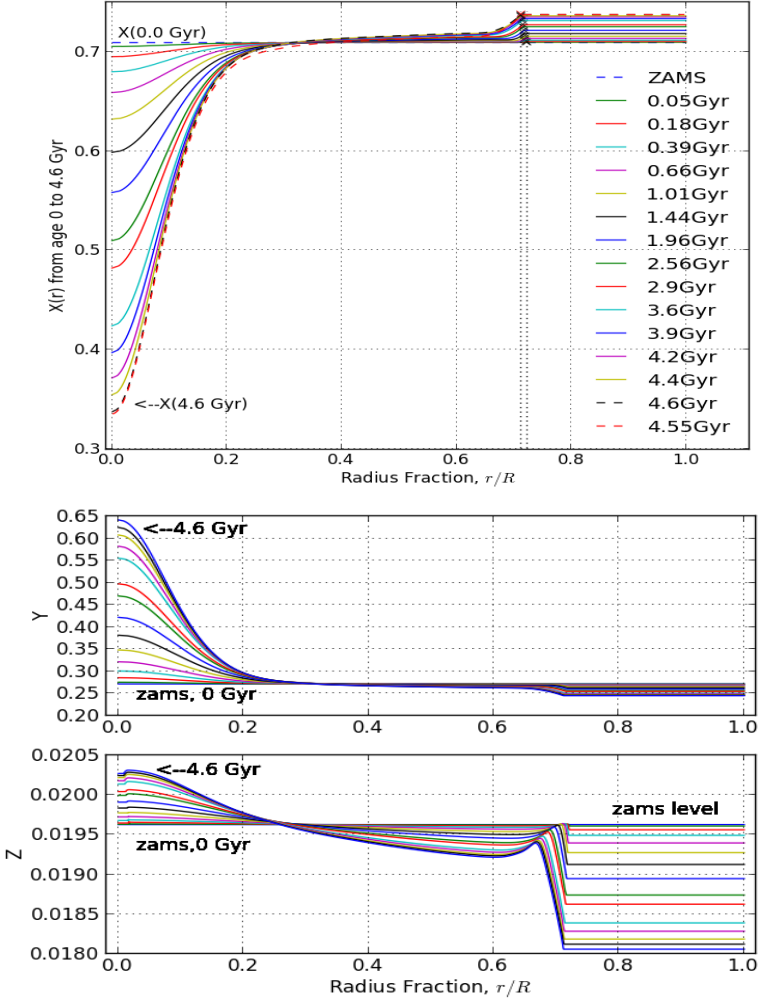

Figure 3. At the top graph, the profile of hydrogen mass fraction $X$ is drawn for all the 15 GONG SSEMs [1] from ZAMS stage (blue dashed line) to present solar age of 4.6 Gyr for GONG SSEM No. 24 (black dashed line). This almost coincides with the profile data from YREC SSEM model No. 741 [3] at age 4.55 Gyr also shown as red dashed line. The lower two graphs show Y and Z profiles (see text).

## 3. Study of the Luminosity Transfer Time Scale in SSEMs

### 3.1 Introduction

The luminosity transfer time from the center of the sun to its photosphere is the sum of photon diffusion time through the RZ up to the tachocline and the convective energy transfer time through the CZ up to the photosphere. The calculation of the photon diffusion time employs the same method adopted by [6]. In [6] the photon diffusion time was calculated in a cruder solar model available by treating the path of photons as a 3-D random walk problem with variable local mean free path ($\ell mfp$). In each shell the $\ell mfp$ has a reciprocal dependence on the product of density and mean opacity. Hence an accurate estimate of the photon diffusion timescale in all the SSEMs is now done here since a state of the art computed Rosseland mean opacity (mass absorption coefficient) and density data are also available for all the grid points. This computation will first be described below starting with a brief description of the discretized expressions used here.

The local mean free path or step-length in a given spherical shell labeled as grid point $i$ is given by,

$$\ell_i = (\kappa_i \rho_i)^{-1} \qquad (3.1)$$

In a given shell of radius $r_i$ and thickness $dr_i \equiv (r_{i+1} - r_i)$ the number $n_i$ of such equal step length vectors required to cross the shell would be quite large since for each step vector all directions are equally probable. In each shell the step length is different in view of equation (3.1). Both the Rossland mass and linear mean absorption coefficients are shown plotted for all the SSEMs together in Figure 4 at the top and bottom respectively along with marking of the tachocline zone (TZ) at the interface between RZ and CZ. At the top it is seen that the opacity gradually rises in the radiative interior up to the TZ mainly due to free-free and bound-free absorption. It begins to rise steeply in the convective zone and becomes very high reaching levels of 100000 in the envelope. The linear opacity decreases slowly until TZ, then increases steeply in the envelope and finally becomes very small in the photosphere allowing a rapid escape of photons (radiation). Therefore the total number, N, of steps a photon requires to reach the radial distance $r_s$ in the sun starting from the center and passing through **s** shells can be written as

$$N_s = \sum_{i=1}^{s} n_i \qquad (3.2)$$

The mean square step length up to shell **s**, $\langle l^2 \rangle_s$ is then given by

$$\langle \ell^2 \rangle_s = \frac{1}{N_s} \sum_{1}^{s} n_i \ell_i^2 \qquad (3.3)$$

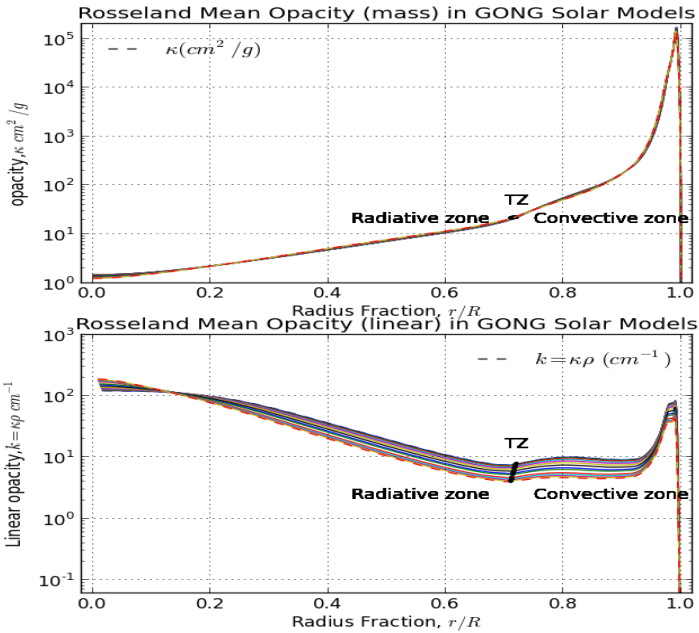

Fig. 4. The Rosseland mean linear absorption coefficient (cm$^{-1}$) in all the 15 solar models is plotted above and the Rosseland mean mass absorption coefficient (opacity, cm$^2$/g) is plotted at the top.

Chandrasekhar [7] showed that mean square step length (3.3) is related to the mean square displacement $\langle r^2 \rangle_s$ by,

$$3 \langle r^2 \rangle_s = N_s \langle \ell^2 \rangle_s = \sum_{1}^{s} n_i \ell_i^2 \qquad (3.4)$$

This expression can now be written, with the displacement $r_{s-1}$ as the radius of previous shell $s-1$, as

$$3\, r_{s-1}^2 = \sum_1^{s-1} n_i \ell_i^2 \qquad (3.5)$$

Then equation (3.4) can then be rewritten as,

$$3\, r_s^2 = \sum_1^s n_i \ell_i^2 = \sum_1^{s-1} n_i \ell_i^2 + n_s \ell_s^2 \qquad (3.6)$$

The number of steps in this shell is obtained by subtracting (3.5) from (3.6) as

$$n_s = 3\,(r_s^2 - r_{s-1}^2)/\ell_s^2 \qquad (3.7)$$

The photon then spends a time $\Delta t_s = n_s\, \ell_s / c$ in this shell where **c** is the speed of light. Then the total time spent in all the **s** shells comprising the RZ in the sun is the radiative diffusion time $t_{rad}$ given by (noting that **s** is a dummy index),

$$t_{rad} = \sum_{i=1}^s \Delta t_i = \frac{3R^2}{c}\sum_{i=1}^s (f_i^2 - f_{i-1}^2)/\ell_i \qquad (3.8)$$

where R is the radius of the SSEM considered and the radius fraction of the shell **s** is defined by $f_s = r_s/R$. The number of shells included in the summation and the radius fraction $f_{TZ}$ of TZ are listed in columns 4 and 5 respectively in Table 1. In this equation the expression under summation has the dimension of inverse length. It represents the reciprocal of an effective mean free path $\langle \ell \rangle_{eff}$ for the entire RZ up to $f_{TZ}$ and is given by,

$$\langle \ell \rangle_{eff} = \left[\sum_{i=1}^{TZ}(f_i^2 - f_{i-1}^2)/\ell_i\right]^{-1} \qquad (3.9)$$

The photon diffusion time given by (3.8) in the SSEM is then also written as,

$$\boldsymbol{t_{rad}} = \frac{3R^2}{c\langle \ell \rangle_{eff}} \qquad (3.10)$$

The maximum value of $\boldsymbol{\ell mfp}$ at TZ, the photon diffusion time needed to reach TZ in each SSEM, the average value $\langle \boldsymbol{\ell mfp} \rangle$ up to TZ, the number of random steps needed to reach TZ and the ratio of luminosity at SSEM age to the present luminosity are respectively listed Table 2. The photon diffusion timescale $\boldsymbol{t_{rad}}$ is plotted against radius fraction depth for all the SSEMs in Figure 5.

Table 2. Listed are SSEM No., age, $f_{TZ}$, $\ell mfp$, $t_{rad}$, $\langle \ell mfp \rangle$, number of steps to reach TZ and the Luminosity ratio.

| SSEM No. | Age Gyr | $f_{TZ}$ $(r/R)_{TZ}$ | $\ell mfp$ cm | $t_{rad}$ $10^5$ yrs | $\langle \ell mfp \rangle$ cm | Steps $\times 10^{25}$ | $L/L_\odot$ |
|---|---|---|---|---|---|---|---|
| 01 | 0.000 | 0.721728 | 0.1288 | 2.045 | 0.0576 | 1.243 | 0.709 |
| 03 | 0.051 | 0.720744 | 0.1293 | 2.043 | 0.0580 | 1.240 | 0.712 |
| 05 | 0.185 | 0.720486 | 0.1311 | 2.039 | 0.0587 | 1.236 | 0.720 |
| 07 | 0.386 | 0.719594 | 0.1347 | 2.031 | 0.0596 | 1.231 | 0.731 |
| 09 | 0.660 | 0.718829 | 0.1396 | 2.022 | 0.0607 | 1.226 | 0.745 |
| 11 | 1.010 | 0.718112 | 0.1462 | 2.009 | 0.0622 | 1.220 | 0.763 |
| 13 | 1.442 | 0.717352 | 0.1550 | 1.994 | 0.0640 | 1.213 | 0.786 |
| 15 | 1.959 | 0.716418 | 0.1666 | 1.975 | 0.0663 | 1.205 | 0.815 |
| 17 | 2.565 | 0.715970 | 0.1813 | 1.953 | 0.0693 | 1.197 | 0.851 |
| 18 | 2.903 | 0.714841 | 0.1909 | 1.939 | 0.0711 | 1.193 | 0.873 |
| 20 | 3.603 | 0.713478 | 0.2127 | 1.911 | 0.0752 | 1.185 | 0.922 |
| 21 | 3.919 | 0.712921 | 0.2237 | 1.899 | 0.0773 | 1.182 | 0.945 |
| 22 | 4.213 | 0.712501 | 0.2344 | 1.886 | 0.0793 | 1.180 | 0.968 |
| 23 | 4.407 | 0.712039 | 0.2422 | 1.878 | 0.0807 | 1.178 | 0.984 |
| 24 | 4.600 | 0.711177 | 0.2503 | 1.870 | 0.0822 | 1.177 | 1.000 |
| 741 | 4.550 | 0.711115 | 0.2540 | 1.883 | 0.0816 | 1.197 | 1.000 |

### 3.2 Discussion of the Radiative Diffusion Time Scale

In the left half of Figure 5 the graphs of X and $\ell mfp$ for each SSEM follow a similar order from age zero at top to 4.6 Gyr last up to a radius fraction depth of about 0.136. In this region the $\ell mfp$ is less than 0.009 cm for the zero age SSEM and it decreases to about 0.005cm at age 4.6 Gyr. Beyond this region a reversal of the order occurs and the details are shown using an expanded scale in the box inset of Figure 5. It is clear that increasing local density due to increase of $\mu(X,Y,Z)$ and opacity causes $\ell mfp$ to decrease in this region. The graphs of X show a similar reversal around a radius fraction depth of 0.4 which is maintained in the right half of Figure 5 all the way up to the model surface along with the corresponding $lmfp$. Here the $\ell mfp$ reaches a local maximum at TZ, then decreases slowly to a minimum at a depth of about 0.97 and finally becomes very large as photons escape from the photosphere. This part of the $\ell mfp$ graph may be compared with the Figure 1 of [6] which shows a double hump feature that is certainly due to the physics behind the solar model used in that paper. This feature is not present in any graph of $\ell mfp$ in Figure 5 including the SSEM No.741 shown by dashed red lines and lying close to that of GONG SSEM at age 4.6 Gyr. This close match suggests a good degree of convergence in the physics and computational methods used in the SSEMs considered.

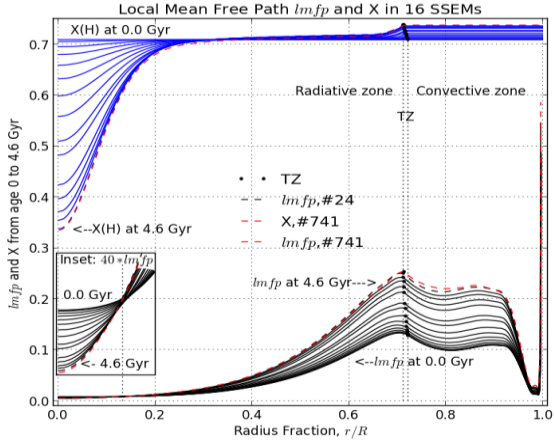

Fig. 5. The local mean free path (***lmfp***) and the hydrogen fraction $X$ are plotted as a function of normalized model radius for each SSEM from age zero to the present.

A graph of corresponding photon diffusion times for the 15 SSEMs is shown in Figure 6. The placement order of each graph in the left side of Figure 6 from top to bottom is just the reverse of that shown in Figure 5. This is evident since photons are increasingly slowed down in the core as evolution progresses and hence the diffusion times also increase until a radius fraction depth of about 0.43 is reached. Beyond this point there is again a reversal of the order as shown in Figure 5 due to the combination of two factors. The continuous increase of $\ell mfp$ right up to TZ and beyond (see Figure 5) again results in decreasing diffusion times in each SSEM. To this must be added the extra increase in $\ell mfp$ and hence an added shortening of diffusion times as evolution progressed. The net result is that the overall photon diffusion timescale needed to reach TZ decreases gradually from its peak value of over 204500 years in the SSEM at zero age to about 187000 years in SSEM at age 4.6 Gyr. The corresponding value of 188300 for SSEM No. 741 at age 4.55 Gyr is also seen to be in good agreement from the listed values in the fifth column of Table 2. Finally both the maximum of $\ell mfp$ at TZ (column 4) and the average mean free path, $\langle \ell mfp \rangle$ (column 6) in Table 2 correlate well with the luminosity ratio of each SSEM in the last column. The photon diffusion time in CZ appears to add a delay of another 70000 to 100000 years but it is neglected here as the convective transfer of energy is far more efficient in transporting luminosity to the photosphere in a matter of a few days as will be shown next. Thus the characteristic timescale for luminosity transport in the sun is the photon diffusion time in RZ.

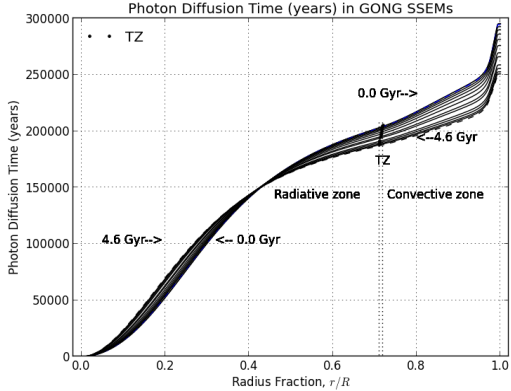

Fig. 6. Here the photon diffusion time in years is plotted for all the SSEMs from age zero to the present. The effective diffusion time is the value computed up to the tachocline zone (TZ) in each model marked by dots at a radius fraction ~0.71. The diffusion time increases in the core region up to a depth of 0.43 as the model age increases and thereafter decreases gradually. The convective zone lies beyond TZ.

## 4. The Convective Energy Transfer Time Scale

The solar convection zone (CZ) has a width of about 29% of the solar radius above the TZ and extends right up to the photosphere. In order to determine the time scale for convective energy transfer the average speed of convective mass flow needs to be known. According to the Schwarzschild criterion convection will occur when the radiative temperature gradient $\nabla_{rad} = (dlnT/dlnP)_{rad}$ exceeds the adiabatic gradient $\nabla_{ad} \equiv (dlnT/dlnP)_{ad} = 0.4$ in any grid point layer. If the actual temperature gradient as determined from model data is defined as $\nabla_T = (dlnT/dlnP)_T$ then the condition $\nabla_{rad} > \nabla_T \geq \nabla_{ad}$ is satisfied in CZ. It is found that a super-adiabatic condition exists in most of CZ (except near the surface) wherein the actual temperature gradient $(dT/dr)_{act} = \nabla_T(dlnP/dr)$ is very slightly larger than the adiabatic temperature gradient $(dT/dr)_{ad} = 0.4T(dlnP/dr)_{ad}$ [9]. The calculation of convective energy transport is currently done using a 60 year old model [9] known as the mixing length theory (MLT).

In the MLT it is assumed that bubbles of gas adiabatically move over a radial distance $\ell_m$ called the mean mixing length and then deliver their excess heat to the surrounding cooler gas. Cyclic convective motions are able to do this continuously and thus very efficiently. The mixing length is generally equated to the local pressure scale height $H_P = |dr/dlnP| = P/\rho g$ which is defined as the radial distance over which the pressure changes by an e-folding factor. In stellar modelling work it is usual to take $\ell_m = \alpha H_P$ where the multiplier parameter is chosen in the range $0.5 \leq \alpha \leq 3$ (GONG SSEMs have $\alpha \approx 2$). In terms of the super-adiabaticity $(\nabla_T - \nabla_{ad})$ the average convective velocity, $V_{con}$ of the convective fluid elements is found [8] to be given by,

$$V_{con} = \sqrt{\alpha \ell_m (\nabla_T - \nabla_{ad})/2} \approx v_s\sqrt{(\nabla_T - \nabla_{ad})} \quad (3.11)$$

where $\alpha = \ell_m/H_P$ determines the multiple of the local pressure scale height selected and $v_s$ is the local speed of sound in CZ which is on the order of $10000 \, m/s$. Equation (3.11) then shows that since $(\nabla_T - \nabla_{ad}) \approx 10^{-5}$, the convective velocity $V_{con} \approx 32 \, m/s$ only. This speed is indeed capable of transporting the solar luminosity through the entire CZ in a matter of less than 75 days and it can be characterized as the convective time scale. Various theoretical estimates of the convective timescale show a range from 20 days [9] to over 70 days [10]. These estimates will now be compared with available solar model calculations.

It was found that stellar model structure data obtainable from EZWeb Tool, an on-demand fast stellar evolution model generating facility [11] contained a listing of convective velocities in the convective zone of a star of selected mass and composition. This EZWeb Tool computes stellar models using a program which can automatically redistribute grid points depending on the age of the model. A set of SSEMs for a solar mass star with $Z = 0.02$ were obtained. Each structure file lists data for just 199 grid points but also provides non-zero convective velocity data computed for about 40 grid points in the CZ. The convective velocity data from EZWeb structure file No. 00065 for sun aged 4.56 Gyr is shown plotted in Figure 7. It may be seen that the flow velocity $V_{con}$ starts from zero at the base of CZ, rises to ~$120 \, m/s$ at $r$~$0.95 \, R_\odot$ and then shoots up to large velocities over $2 \, km/s$ before reaching the photosphere where the flow weakens considerably and hence radiative equilibrium prevails again. A recent estimate [12] of the lower limits to convective flow velocities based on Model S data appears to suggest that the above values are reasonable. These are based on the observed properties of the internal solar differential rotation. After adapting the radial flow velocities to the density profile of Model S provides the following [12] empirical equation:

$$V_{con} \geq \sim 30 \, m \, s^{-1} \left[\frac{\rho}{0.008 \, g \, cm^{-3}}\right]^{-1/2} \left[\frac{r}{0.95 \, R_\odot}\right]^{-1} \quad (3.12)$$

In the above expression the value of $\rho = 0.008 \, g \, cm^{-3}$ is the density at the radius $r$~$0.95 \, R$ in Model S. Thus in the upper CZ at this radius this equation sets the minimum flow velocity to be ~$30 \, m/s$. In the deeper level in CZ at $r$~$0.75 \, R$ where $\rho = 0.14 \, g \, cm^{-3}$ the minimum flow velocity is found to be ~$9 \, m/s$. Since these values are the lower limits, larger flow velocities as shown in Figure 7 are considered probable. Hence the EZWeb SSEM data are used in estimates done in this section.

Since the distance between the grid points is known from the model data the time needed to cross this grid width at an average flow speed can be computed. This is done for all the grid points in CZ and the time in days to cross each grid interval is plotted in Figure 8. This shows that more time is needed for convective flows to cross the first few grid points after starting from the base of CZ. The total time $t_{con}$ required to reach the photosphere from the base of CZ is ~40 days in this case. A similar exercise done with selected

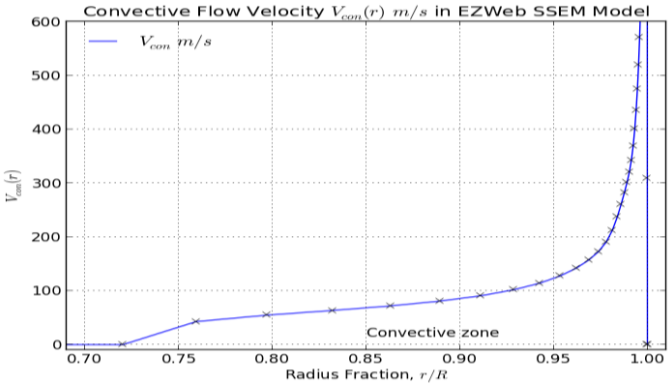

Fig. 7. The convective flow velocity $V_{con}$ in the convective zone of the sun aged 4.56 Gyr .

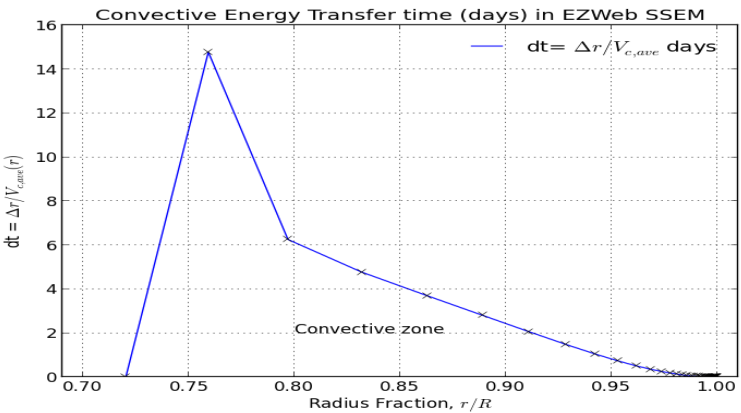

Fig. 8. Plotted is the mean flow velocity in each grid width of about 5000 km showing that a total time of about 40 days is needed for the solar luminosity to be transported across the CZ to the photosphere.

EZWeb SSEM structure files [11] of the solar model from zero age onwards showed similar results. Both $t_{rad}$ and $t_{con}$ as obtained for several EZWeb SSEMs below are listed in Table 3.

Table 3. Time scale for radiative and convective energy transport in EZWeb Solar Evolution Models (Z = 0.02)

| File No. | Age (Gyr) | $t_{rad}(years)$ | $t_{con}(days)$ |
|---|---|---|---|
| 00000 | zero | 196053 | 52.8 |
| 00041 | 0.367 | 195493 | 47.8 |
| 00044 | 0.635 | 193905 | 52.0 |
| 00047 | 1.097 | 194337 | 40.6 |
| 00049 | 1.550 | 191600 | 44.8 |
| 00051 | 1.980 | 188819 | 53.2 |
| 00054 | 2.595 | 187860 | 42.6 |
| 00056 | 2.980 | 185310 | 49.1 |
| 00059 | 3.542 | 184398 | 40.2 |
| 00063 | 4.230 | 179626 | 56.9 |
| 00065 | 4.560 | 180029 | 39.5 |
| 00066 | 4.720 | 178914 | 41.8 |

Both $t_{rad}$ and $t_{con}$ as obtained for several EZWeb SSEMs below are listed in Table 3. This table shows that the definite decreasing trend seen in $t_{rad}$ with model age is in agreement with a similar trend seen in Table 2. However $t_{con}$ (last column) shows no apparent correlation with age of the model. But all these values are well within the range of theoretical estimates stated earlier and lead to an average convective transfer time scale, $\langle t_{con} \rangle$, of about 45 days which can be neglected in comparison with the long radiative transfer time scale shown in the third column of Table 3.

## 4. Time Variation of the Rotational Inertia in SSEMs

### 4.1 Introduction

The fact that each GONG SSEM considered provides mass, density and radius data over 2400 grid points offers a means of accurately calculating the moment of inertia about an axis of rotation (or rotational inertia, MOI) of the sun as it evolved and studying its time variation. Apart from the value of MOI denoted by I, another parameter of interest is the inertia constant, k defined by the expression $I = k\,MR^2$, where M and R are respectively the mass and radius of the sun now (note that $k = 2/3$ is the maximum possible value). A literature search showed that [13] obtained a value of $I = 5.96 \times 10^{46} kgm^2$. This implies that $k = 0.062$ for the sun. A compendium [14] quotes $k = 0.059$. A text book [15] quotes 0.06 and [16] estimates a value 0.062. As these values are based on earlier solar models of uncertain accuracy a fresh determination based on data from SSEMs has been undertaken here.

A brief account of the expressions used for calculating the MOI will now be given. The MOI about an axis of a spherical shell of mass $4\pi r^2 \rho(r) dr$ with radius $r$, density $\rho(r)$ and thickness $dr$ is

$$dI = \frac{2}{3} 4\pi \rho(r) r^4 dr \qquad (4.1)$$

The discretized form of this equation for the $i^{th}$ shell is,

$$dI = \frac{8}{3} \pi \rho_i r_i^4 \Delta r_i = \frac{8}{3} \pi \rho_i r_i^4 [r_{i+1} - r_i] \qquad (4.2)$$

The total MOI is now obtained by summing over all shells as,

$$I = \frac{2}{3} \int_0^R dM(r) r^2 = \sum dI = \sum_{i=0}^{R(i)} \left(\frac{8}{3} \pi \rho_i r_i^4 [r_{i+1} - r_i]\right) \qquad (4.3)$$

This procedure is repeated for all the SSEMs. It was remarked in Section 1 that the bulk of the solar radiative zone (RZ) slowly rotated almost like a rigid body. Hence the MOI of RZ and the inertia constant are also computed for all SSEMS by summing up to the shell ending at TZ.

### 4.2. Age Dependence of the Rotational Inertia in SSEMs

The results of computations are presented in Table 4 which shows the model number, its age, radius, central density, MOI and the inertia constant for all the SSEMs.

Table 4. Listed are the SSEM Number, Age, Central density, total MOI $I_\odot$, Inertia Constant $k_\odot$, MOI $I_{TZ}$ of RZ, and Inertia Constant $k_{TZ}$

| SSEM No. | Age Gyr | $\rho_c$ × $10^3$ kg m$^{-3}$ | $I_\odot$ × $10^{46}$ kg m$^2$ | Inertia Constant $k_\odot = \frac{I_\odot}{M_\odot R_\odot^2}$ | $I_{TZ}$ × $10^{46}$ kg m$^2$ | Inertia Constant $k_{TZ} = \frac{I_{TZ}}{M_{TZ} R_{TZ}^2}$ |
|---|---|---|---|---|---|---|
| 01 | 0.000 | 82.3 | 6.73549 | 0.09112 | 5.76554 | 0.15541 |
| 03 | 0.051 | 82.6 | 6.76489 | 0.09104 | 5.78173 | 0.15451 |
| 05 | 0.185 | 83.7 | 6.80045 | 0.09057 | 5.80954 | 0.15375 |
| 07 | 0.386 | 85.5 | 6.82174 | 0.08983 | 5.82427 | 0.15275 |
| 09 | 0.660 | 88.1 | 6.83754 | 0.08880 | 5.83762 | 0.15127 |
| 11 | 1.010 | 91.7 | 6.85246 | 0.08748 | 5.85260 | 0.14931 |
| 13 | 1.442 | 96.5 | 6.86941 | 0.08586 | 5.87086 | 0.14685 |
| 15 | 1.959 | 103.0 | 6.89114 | 0.08389 | 5.89378 | 0.14386 |
| 17 | 2.565 | 111.6 | 6.92129 | 0.08156 | 5.92867 | 0.14011 |
| 18 | 2.903 | 117.1 | 6.94117 | 0.08025 | 5.94467 | 0.13822 |
| 20 | 3.603 | 130.1 | 6.99174 | 0.07752 | 5.99146 | 0.13397 |
| 21 | 3.919 | 136.9 | 7.01970 | 0.07629 | 6.01686 | 0.13202 |
| 22 | 4.213 | 143.8 | 7.04934 | 0.07513 | 6.04388 | 0.13016 |
| 23 | 4.407 | 148.7 | 7.07086 | 0.07438 | 6.06191 | 0.12898 |
| 24 | 4.600 | 153.9 | 7.09423 | 0.07363 | 6.07863 | 0.12791 |
| 741 | 4.550 | 153.2 | 7.09959 | 0.07369 | 6.07477 | 0.12785 |

In Figure 9 the MOI of each shell $dI(r) = 2/3 r^2 dM(r)$ is plotted against its mass fraction at the top and against the radius fraction below for all the 16 SSEMs. The dominant contribution to the total MOI is seen to arise from shells within about 80% of the model radius at which point 99% of the solar mass is already enclosed. It may be observed that the MOI per shell in the model No. 741 as shown by the dotted line graphs is nearly twice that of all the remaining graphs of 15 SSEMs. This is because of the significantly smaller number of shells for which data are computed in this model. The SSEM No. 741 is described by 1455 shells up to the model surface but only about 580 shells cover its RZ up to the TZ at the bottom of CZ at which point over 97.5% of solar mass is enclosed. For all the SSEMs considered here the corresponding numbers are respectively 2402 and about 1000. The smaller numbers of shells in SSEM No. 741 therefore makes most of them both thicker and more massive. This results in a much larger MOI per shell in that model. But a perusal of the last two rows of Table 4 shows that there is excellent agreement between the values of MOI and $k$ for the two SSEMs listed there, namely model No.24 (Model S) at age 4.60 Gyr and model No.741 at age 4.55 Gyr respectively. In both parts of Figure 9 the effect of evolution on each SSEM can be seen by the gradual leftward shift of the corresponding graph from zero-age to 4.6 Gyr. The corresponding SSEM data indicate that during this period the central density has increased from about $82000 \text{ kg m}^{-3}$ to $154000 \text{ kg m}^{-3}$ as shown in the third column of Table 4. Thus there is a gradual increase in the core mass distribution but a negligible increase in the corresponding MOI. Since this is coupled with a slow expansion of the corresponding SSEM as shown in the third column in Table 1 there is only a small gradual increase in the total MOI of the model from $6.73 \times 10^{46}$ kg m² at zero age to $7.09 \times 10^{46}$ kg m² at 4.6 Gyr but the factor $k$ decreases from 0.0911 to 0.0736 during the same period.

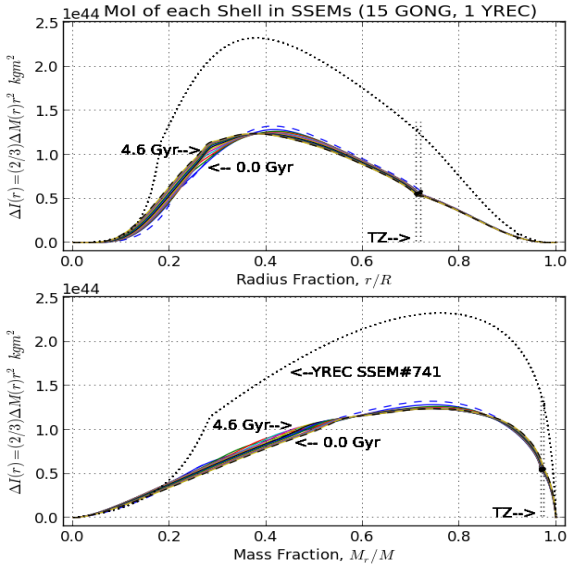

Figure. 9. The MOI of each shell is plotted against its mass fraction above and against the radius fraction at the top for all the SSEMs. Notice the leftward shift of the graphs with age. Notice that mass distribution in the region $0.2 < r/R < 0.71$ or $\sim 0.4 < M_r/M < 0.98$ dominates the total MOI.

The MOI of the nearly rigid RZ and the corresponding inertia constant factor have also been calculated taking into account the radius and the total mass of this sphere. These quantities have been listed in the last two columns of Table 4. It is now seen that MOI of RZ increases from $5.76 \times 10^{46}$ kg m² at zero age to $6.08 \times 10^{46}$ kg m² at 4.6 Gyr and the factor $k$ decreases from 0.1554 to 0.1279 during the same period.

### 4.3. Discussion of the Internal Changes in MOI

An investigation of how the mass distribution in different parts of the solar radiative interior contributes to the total MOI of the SSEM was made by plotting the spherical shell mass $dM(r) = 4\pi r^2 dr\rho(r)$ and the corresponding MOI $dI(r) = 2/3 r^2 dM(r)$ after normalizing them as shown in Figure 10. Computations show that

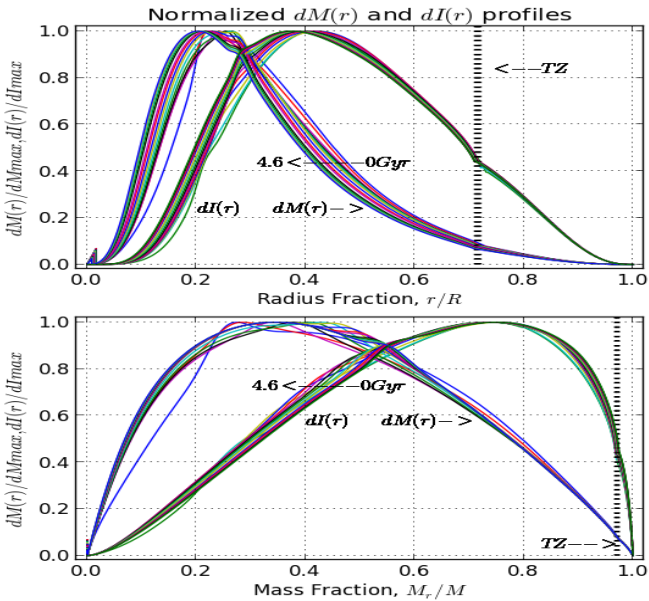

Figure. 10. Both the normalized mass of each shell $dM(r)$ and its MOI $dI(r)$ are plotted. Notice that the bunched curves show a shift of the 0 Gyr curve at right to the 4.6 Gyr curve at left.

the region contributing over 80% to the total solar MOI can be nearly identified as the mass distribution in the range $0.2 < r/R < 0.71$ or $\sim 0.4 < M_r/M < 0.98$ in all the SSEMs. Hence the entire RZ was split into three parts. The first part is the central core region $0 < r/R < 0.2$, the second part is the outer core region $0.2 < r/R < 0.4$ and the third part is the remaining region $0.4 < r/R < 0.71$ which ends with TZ at the base of the convective zone. The MOI contribution from these three parts denoted as $I(0 < r/R < 0.2)$, $I(0.2 < r/R < 0.4)$ and $I(0.4 < r/R < 0.71)$ respectively were computed separately in all the SSEMs. The sum of these three MOIs would of course be equal to that of entire radiative interior given by $I_{TZ}(0 < r/R < 0.71)$. The mass content of these regions was also computed. The results of these computations have been grouped together for the entire sun and shown in Table 5. The initial values

| Table 5. Region wise fractional Moment of Inertia and mass content in the SSEMs | | | |
|---|---|---|---|
| **Region** | **Value at 0 Gyr** | **Parameter** | **Value at 4.6 Gyr** |
| | $609615 \ km \leq$ | Radius, $R_\odot$ | $\leq 695978 \ km$ |
| | $6.73549 \times 10^{46} \ kgm^2 \leq$ | MOI, $I_\odot$ | $\leq 7.09423 \times 10^{46} kgm^2$ |
| **Region 1 (central core):** | $0.037 \cdot I_\odot \leq$ | $I\left(0.0 < \frac{r}{R} < 0.2\right)$ | $\leq 0.062 \cdot I_\odot$ |
| Mass content : | $0.218 \cdot M_\odot \leq$ | $M\left(0.0 < \frac{r}{R} < 0.2\right)$ | $\leq 0.342 \cdot M_\odot$ |
| **Region 2 (outer core) :** | $0.322 \cdot I_\odot \leq$ | $I\left(0.2 < \frac{r}{R} < 0.4\right)$ | $\leq 0.349 \cdot I_\odot$ |
| Mass content : | $0.497 \cdot M_\odot \leq$ | $M\left(0.2 < \frac{r}{R} < 0.4\right)$ | $\leq 0.453 \cdot M_\odot$ |
| **Region 3 (up to TZ) :** | $0.498 \cdot I_\odot \leq$ | $I\left(0.4 < \frac{r}{R} < 0.71\right)$ | $\leq 0.447 \cdot I_\odot$ |
| Mass content : | $0.257 \cdot M_\odot \leq$ | $M\left(0.4 < \frac{r}{R} < 0.71\right)$ | $\leq 0.181 \cdot M_\odot$ |
| **Radiative Zone** : | $0.857 \cdot I_\odot \leq$ | $I_{TZ}\left(0.0 < \frac{r}{R} < 0.71\right)$ | $\leq 0.858 \cdot I_\odot$ |
| (total of all 3 regions) : | $0.972 \cdot M_\odot \leq$ | $M\left(0.0 < \frac{r}{R} < 0.71\right)$ | $\leq 0.976 \cdot M_\odot$ |
| **Convective Zone** : | $0.144 \cdot I_\odot \leq$ | $I\left(0.71 < \frac{r}{R} < 1.0\right)$ | $\leq 0.143 \cdot I_\odot$ |
| Mass content : | $0.028 \cdot M_\odot \leq$ | $M\left(0.71 < \frac{r}{R} < 1.0\right)$ | $\leq 0.024 \cdot M_\odot$ |

corresponding to the sun at zero age are shown at the left side of all ranges and the current values corresponding to the sun at age 4.6 Gyr are shown at right. An inspection of entries in Table 5 leads to the following observations. The dominant contributor to the total MOI at any age is the mass distribution in the outer core region (Region 2) together with mass distribution up to the TZ (Region 3). For the sun at zero age it is seen that the total mass content in these two regions of 75.4 per cent gives rise to 82 percent of the total MOI of the sun as it was then. For the evolved sun at age 4.6 Gyr the mass content of regions 2 and 3 drops to 63.4 percent but still it contributes as much as 80 percent of total MOI due to expansion of the sun by 14 percent. This finding may be contrasted with the suggestion by [13] that 80 per cent of solar MOI now is generated by the mass distribution between radius fractions 0.20 and 0.65 only. The 12 per cent mass content that disappeared from these two regions during the solar evolution until now appears to have moved into in the central core region (Region 1) which now has a mass content of 34.2 percent while it was just 21.8 per cent at zero age of the sun.

However, the contribution of the mass content of the central core (Region 1) to the total MOI of the sun is quite small. While it was just 3.7 per cent at zero age it has now risen to 6.2 percent in spite of the mass content being over one third of solar mass now. The main reason for the central core MOI being a small fraction of the total solar MOI is that the range of radii involved is relatively small even though the core radius has now increased to 140000 km from about 122000 km at zero age. It is in the central part of this region that a severe depletion of hydrogen content (X) and its replacement by helium (Y) has occurred. This can be seen in Figure 3 which shows the age dependent profiles of X, Y and Z for all the 15 GONG SSEMs. Some increase in the mass build up of the central core has also been due to loss of hydrogen from there by buoyant diffusion towards the convective zone and a small gain in terms of diffusive settling of helium and metals from the outer radiative zone and the convective zone of the sun. This is suggested by the fact that the current levels at the center are X = 0.33727, Y = 0.64246 and Z = 0.02027 whereas in CZ they are X = 0.73729, Y = 0.24465 and Z = 0.01806. As a result of these factors the central density in the sun has risen by about 90 per cent over its zero age value.

Thus it is seen that the entire RZ in the sun contributes about 85.7% of the total solar MOI at zero age and it surprisingly remains almost the same even now in spite of changes in the interior regions. The remaining 14.3% of the solar MOI is accounted for by the convective zone and it too remains nearly unchanged during the long solar evolution. This happens because of the following circumstances. At zero age of the sun the CZ had a mass content of just 2.8% of the solar mass but it stretched over an expanse of about 177000 km between radii 433000 km and 609615 km. In all the SSEMs this region is described by data for about 1400 grid points (shells) having an average thickness of 140 km. At the current age of the sun the CZ mass content got reduced to 2.4% but its extent increased to over 195000 km between radii 501000 km and 696000 km due to expansion of the sun. These circumstances have combined to keep the MOI of the convective zone at nearly constant levels as shown in Table 4.

Another parameter of interest is the variation of the solar MOI with model age from its value at zero age of the sun. This is found (see Table 4) to be a modest *increase* at a rate of about 1.175% per Gyr or about $2.47 \times 10^{28} \ kg \ m^2 s^{-1}$. This finding contradicts the observation by [13] that the solar MOI *decreases* at the rate of $5.5 \times 10^{27} \ kg \ m^2 s^{-1}$.

The angular momentum of the radiative zone of the sun is also of interest since it was pointed out earlier (Section 1) that this region rotates like a near rigid body. If the angular velocity is taken to be $\Omega = 2\pi \times 433 \times 10^{-9} \ s^{-1}$ and the current value of the MOI of solar RZ is taken as $I = 6.08 \times 10^{46}$ kg m² then the corresponding angular momentum is given by $I\Omega = 1.65 \times 10^{41} \ kg \ m^2 \ s^{-1}$.

## 5. Conclusions

The work described in Section 2 leads to the following conclusions. The extent of solar radiative zone has been determined (Table 1) to an accuracy of at least one shell thickness below the base of the convection zone in all the SSEMs. It was found that both the relevant mass **fraction** and the radius **fraction** were essentially constant during solar evolution up to the present. It can therefore be concluded that the fractional extent of radiative and convective zones in a well constructed model of a solar mass star are fixed in the ZAMS stage itself and that the extent of these regions remains essentially unchanged during a significant part of its main sequence lifetime. This follows from the observation that while the spread of the shell burning region in the sun has caused some expansion it still remains too far away to affect the tachocline zone. More detailed standard solar models are needed to understand how the step up of hydrogen as well as step down of helium and metals occur at the tachocline zone.

The work described in Section 3 leads to the following conclusions. In the central nuclear burning region of the sun the photon $\ell mfp$ has the smallest possible value in the entire RZ and so the photons diffuse out slowest there. The $\ell mfp$ keeps decreasing with model age there as it is sensitive to changing central conditions. Both the average photon mean free path and its maximum value at the tachocline zone are seen to be important characteristics of an SSEM of a given age as they are well correlated with the corresponding luminosity. These two parameters are found to increase from 0.0576 cm to 0.0822 cm and from 0.1288 cm to 0.2503 cm respectively during the solar evolution up to the present. This is why the solar photon diffusion timescale has gradually decreased from 205000 years at zero age to 187000 years now. The mean value of the photon diffusion timescale at 196000 years, however, is a characteristic of the size of the solar radiative zone which is set by the mass and chemical composition of the sun.

On the other hand the time scale for convective energy transfer in the solar CZ is found to range between 37 and 53 days with an average value of about 45 days. Whatever this may be it is so short and fast compared with the photon diffusion time scale in the sun. Therefore it may be asserted that the photon diffusion time scale of a solar mass star is also the characteristic luminosity transport time scale.

The work described in Section 4 leads to the following conclusions. A perspective on the evolution of the sun through changes in its internal moment of inertia has been obtained. The rotational MOI of the sun in its present state of evolution is found on the basis of the 2402 grid point data of the Model S to be $7.09 \times 10^{46}$ kg m². The corresponding inertia constant is 0.0736 indicating that the sun is highly centrally condensed. This MOI value is nearly 18% larger than the available published values (Section 4.1). Thus the value of the solar MOI determined from data of Model S is the best value available now for the sun. The average rate of increase of MOI during the solar evolution has been about 1.175 % per Gyr or about $2.47 \times 10^{28}\ kg\ m^2 s^{-1}$. The MOI of the almost rigid solar RZ itself now and the inertia constant are $6.08 \times 10^{46}$ kg m² and 0.1279 respectively.

## 6. Acknowledgements

I wish to thank J. C-Dalsgaard of Aarhus University, Denmark for making available for this study 15 SSEM data files of GONG project from age zero onwards to the present, D. B. Guenther of Saint Mary's University, Canada for allowing download of his YREC_d3_7d_grey (circa 2010) SSEM data file and R. Townsend, University of Wisconsin, Madison, USA for making available the use of his EZ Web Tool facility from their respective homepages. I also wish to thank my former mentor J. B. Tatum of the University of Victoria, Canada for sending his extensive lecture notes on computing the moment of inertia of large spheres and for his witty remarks stressing that for usage in rotational kinematics the moment of inertia of any object must always be explicitly referenced to an axis but it is of only a theoretical significance when referenced to a point. The processing of 16 large model data files, computational work, data mining and plotting have been done by writing Python scripts and using the Pythonxy 2.7.5 software.

## References

[1] J. Christensen-Dalsgaard et al., "The Current State of Solar Modeling", *Science*, vol. 272, pp. 1286-1292, 1996.

Homepage: http://astro.phys.au.dk/~jcd/solar_models/

[2] J. C-Dalsgaard, http://owww.phys.au.dk/~jcd/solar_models/sequence/ 2009

[3] P. Demarque, D. B. Guenther, L. H. Li, A. Mazumdar and C. W. Straka, "YREC: the Yale rotating stellar evolution Code", *Astrophys Space Sci*, vol. 316, pp. 31-41, 2008. http://www.ap.smu.ca/~guenther/ evolution/ssm2010.

[4] R. Howe, "Solar Interior Rotation and its Variation", *Living Rev. Solar Phys.*, vol. 6, pp.1-75; http:// www. livingreviews.org/lrsp-2009-1, 2009

[5] J. C-Dalsgaard and M.J. Thompson, "Observational Results and issues concerning the Tachocline", in The Solar Tachocline, Ed by R. Rosner and N. Weiss, Cambridge University Press, Cambridge, 2007.

[6] R. Mitalas and K. R. Sills, "On the Photon Diffusion Time Scale for the Sun", *Astrophys. J.*, vol. 401, 759-760, 1992.

[7] S. Chandrasekhar, "Stochastic Problems in Physics and Astronomy", *Rev. Mod. Phys*, vol. 15, pp. 1-89, 1943.

[8] B. W. Carroll and O. J. Ostlie, "An Introduction to Modern Astrophysics", Addison-Wesley, 1996.

[9] M. Schwarzschild, "Structure and Evolution of the Stars", Princeton University Press, USA, 1958.

[10] J. C-Dalsgaard, "Lecture Notes on Stellar Structure and Evolution", Homepage (see [1]), 2008.

[11] R. Townsend, www.astro.wisc.edu/~townsend/tools, 2013.

[12] M. S. Miesch et al., "On the Amplitude of Convective velocities in the Deep Solar Interior", ArXiv: 1205.1530v2 [astro-ph.SR], 2012.

[13] K. H. Schatten, "Time Variations of the Angular Momentum of the Sun", *Ap. J.*, vol. 216, pp. 650-653, 1977.

[14] A. N. Cox, "Allen's Astrophysical Quantities", Springer, 2000.

[15] W. K. Hartmann, "Moons & Planets", Cengage Learning, 2005.

[16] E. Mamajek, "Basic Astronomical Data for the Sun (BADS)", www.pas.rochester.edu/ ~emamajek / (Notes on the Sun), 2014.

[17] M. N. Anandaram, "Emden's Polytropes", Mapana J Sci, vol.12, pp. 90-114, 2013.